\documentclass[aps,prl, twocolumn,amsmath,amssymb, superscriptaddress,citeautoscript,floatfix]{revtex4-2}
\usepackage{amsbsy}
\usepackage{wasysym}
\usepackage{bm}
\usepackage{color}
\usepackage{textcomp}
\usepackage{tikz-feynman,contour}
\tikzfeynmanset{compat=1.1.0}
\tikzfeynmanset{/tikzfeynman/momentum/arrow shorten = 0.3}
\tikzfeynmanset{/tikzfeynman/warn luatex = false}
\usepackage{gensymb}
\usepackage[breaklinks,colorlinks = true,linkcolor = red,urlcolor=cyan,citecolor=red]{hyperref}
\usepackage{array}
\usepackage{float}
\usepackage{graphicx}
\usepackage{subfigure}
\usepackage{float}
\usepackage{soul}

\usepackage{comment}
\usepackage{dcolumn,multirow}

\begin{document}

\title{%
Chiral spin liquid instability of the Kitaev honeycomb model with crystallographic defects
}

\author{Arnab Seth}
\affiliation{School of Physics, Georgia Institute of Technology, Atlanta, GA 30332, USA}
\author{Fay Borhani}
\affiliation{School of Physics, Georgia Institute of Technology, Atlanta, GA 30332, USA}
\author{Itamar Kimchi}
\affiliation{School of Physics, Georgia Institute of Technology, Atlanta, GA 30332, USA}

\date{February 23, 2026}

\begin{abstract}
We study the spin-1/2 Kitaev honeycomb gapless spin liquid in the presence of Stone-Wales-type local lattice defects with odd-sided plaquettes. While the clean Kitaev model has no finite-temperature phase transitions, we find that introducing a finite defect density $n_d\approx 10^{-4}$--$10^{-2}$ produces a true phase transition with a sizeable $T_c \approx 2 n_d$ in units of the Kitaev exchange. The resulting non-Abelian chiral quantum spin liquid exhibits scalar spin chirality and electron orbital magnetization which peak near lattice defects. This disorder-driven instability relies on an emergent long range ferromagnetic interaction $r^{-\gamma}$ ($\gamma \approx 2.7$) between defect chiralities, mediated by the nearly-gapless fermions, with implications for topology generation in Dirac cones with fluctuating mass terms.  %
\end{abstract}

\maketitle

Crystallographic defects are always present in solid state materials such as magnetic insulators but their role in the highly entangled quantum spin liquid (QSL) phases of frustrated quantum magnets is not well understood. 
Theoretical efforts 
\cite{willans_disorder_2010,willans_site_2011,sreejith_vacancies_2016, nasu_thermodynamic_2020,kao_vacancy_2023,dhochak_magnetic_2010,kao_disorder_2021,kao_vacancyinduced_2021,singhania_disorder_2023,zschocke_physical_2015,knolle_bonddisordered_2019,
cassella_exact_2023, grushin_amorphous_2023,petrova_unpaired_2014,
sanyal_emergent_2021,halasz_doping_2014,halasz_coherent_2016,
lahtinen_perturbed_2014, 
udagawa_visonmajorana_2018,
kimchi_valence_2018,kimchi_scaling_2018,
otten_dynamical_2019, nasu_spin_2021,freitas_gapless_2022,dantas_disorder_2022,
song_lowenergy_2016,yatsuta_vacancies_2023,vojta_kondo_2016}
have focused on  
the paradigmatic 2D Kitaev honeycomb QSL \cite{kitaev_anyons_2006}  which is theoretically solvable and offers promising material realizations 
\cite{jackeli_mott_2009, savary_quantum_2016,
hermanns_physics_2018, takagi_concept_2019, trebst_kitaev_2022,  matsuda_kitaev_2025}. This spin-1/2 model's low energy description has the electrons fractionalizing into gapless Majorana fermions coupled to an emergent gauge field.  However, these low energy features have not yet been unambigously detected in any of the material candidates, suggesting that additional features of the real materials, including crystal defects, might modify the theoretically predicted observables in qualitatively new ways.

It was known since Kitaev's original paper \cite{kitaev_anyons_2006}  that plaquettes with an odd number of sides  generate time reversal breaking terms in the honeycomb model.
Odd-sided plaquettes have therefore been studied in various decorated lattices such as the ``star" lattice of the Yao-Kivelson model \cite{yao_exact_2007, dornellas_kitaevheisenberg_2024}
and pentaheptite lattice \cite{peri_nonabelian_2020}.
Recently they have also been studied on amorphous and polycrystalline lattices \cite{cassella_exact_2023, grushin_amorphous_2023}.
The time reversal breaking in these models produces a chiral spin liquid phase, a gapped QSL whose signatures are distinct from those of the gapless Kitaev QSL, and whose emergent non-Abelian gauge theory is also relevant for topological protection of quantum information processing. 

Three aspects of this prior literature limit its direct applicability to the experimentally relevant case of a Kitaev honeycomb material with crystallographic defects. (1) First is the density of odd sided plaquettes. Most prior work considered only an order-one density, which is incompatible with the view of a honeycomb lattice material with typical dilute density of crystallographic defects. 
Ref.~\cite{grushin_amorphous_2023}  studied polycrystalline lattices in a Voronoi construction that enables a tunable fraction of generic odd sided plaquettes, though only down to a fraction of 0.05. Ref.~\cite{petrova_unpaired_2014} considered 5-7 dislocation defects but focused on the trivially gapped Kitaev model. (An extension to dislocations in the gapless Kitaev model was considered in Ref.~\cite{borhani_realspace_2025}.)
(2) Second is the qualitative nature of the defects:  the previously considered cases involved topological crystalline defects such as dislocations \cite{petrova_unpaired_2014, borhani_realspace_2025} or, in polycrystalline case, effectively disclinations \cite{grushin_amorphous_2023}. In contrast to local (plastic) defects, the topological defects are not locally creatable. Thus an isolated defect necessarily involves an infinite string of switched Kitaev bond labels \cite{borhani_realspace_2025}. This is an unrealistic scenario for magnetic insulators, which brings us to the third consideration: (3) the in-principle realizability of Kitaev bond label  3-coloring with defects.   
Kitaev's $x,y,z$ spin interactions can arise in spin orbit coupled magnetic insulators through the Jackeli-Khaliullin mechanism of edge sharing octahedra. This mechanism has not been previously considered in the presence of a lattice defect.
Even before considering chemical or energetic factors for any particular material candidate, 
insisting on in-principle realizability of 3-colored Kitaev interactions provides geometrical  considerations beyond graph theory. 
In addition to other conditions, these considerations require a \textit{local} crystallographic defect.

\paragraph{\textit{Stone-Wales defects ---}}

\begin{figure}[t]
    \centering
    \includegraphics[width=1\columnwidth]{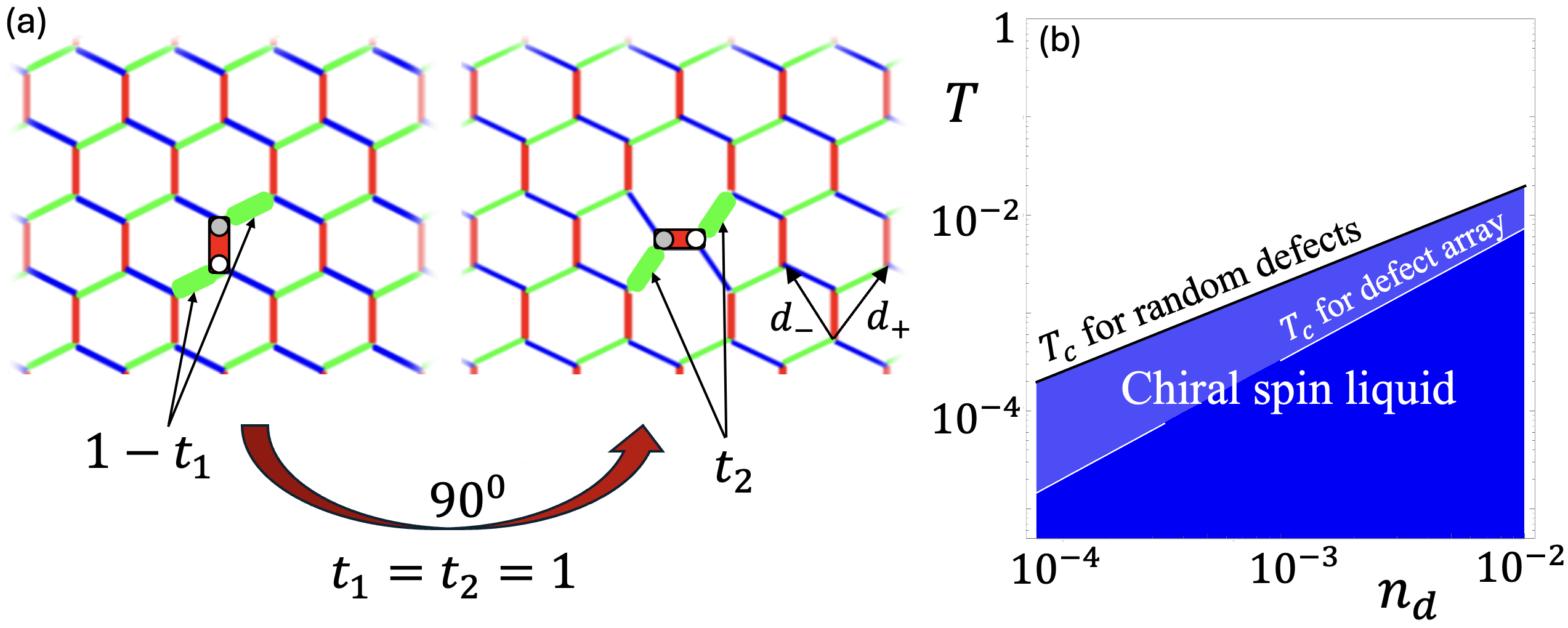}
    \caption{{\bf Local crystallographic defects generating a chiral spin liquid instability.} (a) Stone-Wales (SW) defect as a local 90$^\circ$  bond rotation preserving Kitaev 3-coloring.
    The rearranged connectivity following the bond rotation can be captured by modifying bond strengths via $t_1{=}t_2{=}1$ as shown. 
    This enables perturbative treatments (Eq.~\ref{eq_sw_real space}) whose resummation agrees with numerics (Fig.~\ref{fig2}). 
    (b) Resulting phase diagram with temperature $T$ (in units of  Kitaev exchange $J_K$) and defect density per site $n_d$. 
    Spatially random defects generate a finite temperature time-reversal-breaking phase transition with $T_c\approx 2 n_d$ (black line). 
    Spatial correlation of defects (e.g.\ array) reduces $T_c$ (white line) \cite{seth_generation_2025}.  Below $T_c$ the defects produce nonzero net chirality and lead to a chiral spin liquid. 
     }
    \label{fig1}
\end{figure}

In this work we investigate the case of dilute and realizable odd-sided-plaquette \textit{local} defects in Kitaev honeycomb materials. We do so by defining and analyzing a Kitaev model variant of the so-called \textit{Stone-Wales} (SW) defect well known  from graphene \cite{stone_theoretical_1986,vozmediano_gauge_2010, kot_band_2020}. 
This SW defect consists of a  $\pi/2$ rotation of a honeycomb lattice bond (Fig.~\ref{fig1}) that produces two pentagon-heptagon pairs. The resulting crystallographic defect can also be considered as a   fundamental building block for locally-creatable (non-topological) defects that is a bound state of two 5-7 dislocations with opposite Burgers vectors  or equivalently a quadrupole of 5 and 7 sided disclinations.
Within  the Jackeli-Khaliullin mechanism for generating Kitaev spin exchanges in Mott insulators\cite{jackeli_mott_2009},  certain $\pi/2$ rotations of a hexagonal bond exactly preserve Kitaev bond labels. 
The ideal rotations involve a cubic (0,0,1) axis tilted from the (1,1,1) honeycomb plane, which enables a SW defect with localized out-of-plane deformations; these geometric considerations are discussed in a companion paper \cite{seth_generation_2025}.
In the idealized limit, the SW defect produces an exactly solvable modified  Kitaev model defined %
on a graph consisting of the honeycomb lattice modified by the 3-colored SW defect (Fig.~\ref{fig1}). The local geometric modifications produce additional perturbation terms near the defect.   
A broad class of realistic local crystalloraphic defects in Kitaev honeycomb materials can thus be considered starting from the SW limit.

The resulting spin Hamiltonian in presence of SW defects at locations $r$ can be written as
\begin{align}
    H=J_K\sum_{\langle ij\rangle}\sigma^{\alpha_{ij}}_i\sigma^{\alpha_{ij}}_{j}+\sum_{r} \delta H_{r}
    \label{eq_kitaev}
\end{align} 
with Pauli matrices  $\sigma_j$ here denoting the spin-1/2 at site $j$. The first term is the modified-connectivity Kitaev model with bond ``color'' labels $\alpha_{ij} \in (x,y,z)$ set by the modified graph shown in Fig.~\ref{fig1} for each defect location.
The second term $\delta H_{r}$ consists of the perturbations due to the defects beyond the rearranged connectivity. Importantly, these perturbations are nonzero only in a local region around the defect at location $r$ (possibly decaying quasi-locally for the case of strain fields). 
In the rest of this paper we set $J_K=1$ and study the Hamiltonian $H$ in the limit $\delta H_{r}\rightarrow 0$. %
The results are expected to generalize to any local crystallographic defect with odd sided plaquettes (albeit with modified $\gamma,r_0$, see below). 
The observed gap of the chiral QSL  is expected to provide stability for sufficiently small  $\delta H_{r}$. 
In a companion paper \cite{seth_generation_2025} we extend the numerical study to several solvability-preserving perturbations $\delta H_{r}$, identifying a window of stability for the chiral QSL and computing the modified $\gamma,r_0$ parameters, thereby confirming that the qualitative conclusions below remain robust.

\paragraph{\textit{Defect Fluxes  --- }}

Solving Eqn. \ref{eq_kitaev} with Kitaev's now-standard approach \cite{kitaev_anyons_2006} produces a quadratic Majorana fermion model coupled to a discrete gauge field. 
On hexagonal plaquettes the gauge field produces $0$ or $\pi$ fluxes with corresponding phase accumulation $\pm 1$ by the Majorana fermions; the fermion energy is minimized by the zero flux sector.  
On the odd sided plaquettes the gauge field  produces $\pm \pi/2$ fluxes whose phases are correspondingly imaginary $\pm i$. 
To determine the energetically preferred fluxes on the odd plaquettes we numerically diagonalized systems with one or multiple SW defects with various flux configurations, including the cases of two SW defects with net $\pi$-flux bound to each, involving a nonlocal gauge field. 

We find that each SW defect has a pair of degenerate ground state flux configurations related to each other by time reversal.  
These two states, which consist of $\pm \pi/2$ ($\mp \pi/2$) fluxes on 5 (7) sided plaquettes, can be labeled by an Ising variable $\mu^z_r=\pm 1$ with $r$ denoting the defect location.
We denote the $\mu^z=\pm 1$ pair of ground states as ``Lieb-flux"  states since their property of opposite fluxes on 5 and 7 sided plaquettes corresponds to an (unproven) extension of Lieb's theorem \cite{lieb_flux_1994}, which is consistent with numerical results on the pentaheptite lattice \cite{peri_nonabelian_2020}, amorphous lattices \cite{cassella_exact_2023, grushin_amorphous_2023}, and 5-7 dislocations \cite{petrova_unpaired_2014,borhani_realspace_2025}. 
Analytical justification for Lieb-flux states being the ground states can also be obtained in a different limit, that of the bond anisotropic trivially gapped Kitaev ``A" phase, as we discuss elsewhere \cite{seth_generation_2025}. The study of the excited state flux configurations, their energetics and local contributions to chirality (including an unusual combined monopole-quadrupole excitation), is also reserved for a separate discussion \cite{seth_generation_2025}.

\paragraph{\textit{Topological Majorana Gap and Chern Number --- }}

The locally creatable nature of SW defects means they can be mathematically captured by  a sum of local terms $V$ added to the usual honeycomb Majorana fermion model $H_0$. 
Using this picture we can analyze the effects of SW defects on the Majorana band structure by computing the self consistent T-matrix and its projection to the Majorana fermion low energy Dirac cones. 
As we shall now show,  we find that the SW defects generate both trivial and topological mass terms, and that (with the caveat of several subtleties) the topological mass term generally dominates, generating a Chern number $C=\mu^z=\pm 1$ and associated chirality, which we also verify numerically.

Recall \cite{kitaev_anyons_2006} that $H_0$ can be written (in the standard gauge choice for the zero-flux ground state sector) as
$H_0=\sum_{R,\nu}  ic_{R,A}c_{R-d_\nu,B}$.
The sum is over unit cells $R$, each of which has three bonds, labeled by (and summed over) $\nu=0,+,-$, and associated with a Bravais lattice vector $d_\pm=\frac{\sqrt{3}}{2}\left(\pm 1, \sqrt{3}\right)$ (Fig.~\ref{fig1}) or with $d_0=0$. 
 There is one Majorana fermion
 ($c=c^\dagger$ with $c^2=1$) at each unit cell $R$ and sublattice $A,B$.  
Note that each term  is equal to its Hermitian conjugate 
$-i c_{R-d_\nu,B}c_{R,A}$. 

In the presence of a SW defect in the Lieb-flux $\mu^z=\pm 1$ state, the effective Hamiltonian for Majorana fermions is modified by the addition of the following operator:
\begin{align}
&V^\text{SW}_\text{Lieb}= \left(c_{R,A} , c_{R,B}\right)\left(
t_1 \sigma^y - i \nu t_2\mu^z \sigma^0
\right)\left(
    \begin{array}{c}
         c_{R+d_\nu,A}  \\
         c_{R-d_\nu,B} 
    \end{array}
    \right)
\label{eq_sw_real space}
\end{align}
This corresponds to a SW defect created by rotating the bond with midpoint $R$ by a  $\nu \pi/2$ rotation with $\nu=\pm1$.  Figure \ref{fig1} shows a $\nu=+1$ rotation. 
Here 
$\sigma$ are Pauli matrices acting on sublattice   and 
$\sigma^0$ is the identity matrix.
The 3-colored SW defect becomes fully formed at $t_1=t_2=1$. 
\footnote{This limit is not preserved by the low energy projection, requiring us to keep track of the sense of rotation $\nu$.} 
Observe that while the $t_1$ term merely subtracts a corresponding term from $H_0$, the nonbipartite $t_2$ term breaks time reversal.

To investigate the effects of the SW on the Majorana band structure we can project $V^{\rm SW}_{\rm Lieb}$ into the low energy theory $P$ of the clean model $H_0$. This theory consists of two Majorana Dirac cones,      $P[H_0] = v_F\sum_{ q} {\psi}^{\dagger}_{ q}\left(q_y\sigma^x-q_x\tau^z\sigma^y\right){\psi}_{ q}$ with Fermi velocity $v_F=3/2$.
$\sigma$ and $\tau$ are Pauli matrices acting on sublattice and valley ($K,K'$) respectively. Similarly projecting $V$ we obtain
\begin{align}
&P[V_\text{Lieb}^\text{SW}]=\frac{1}{\mathcal{N}_c}\psi^\dagger_q\left(
\sqrt{3} t_2 \mu^z\tau^z\sigma^z
-t_1    \vec{m}_\text{t}{\cdot}\vec{\tau}_{x,y}\sigma^y\right)\psi_q
\label{eq_sw perturbation}
\end{align}
where
${\psi}^{\dagger}_{ q}=\left(c_{{ K+q},A}~~c_{{ K+q},B}~~c_{{ K'+q},A}~~c_{{ K'+q},B}\right)$. %
$\mathcal{N}_c$ denotes the  number of unit cells.
An additional $\nu$ dependent term involving $\sigma^y$ and $\sigma^x\tau^z$ also appears but merely shifts the Dirac cones in an inversion symmetric manner and can be ignored here (see End Matter). 
Our focus is the two competing mass terms, $t_1$ and $t_2$, which commute with each other but  anticommute with $P[H_0]$. 
The $t_1$ terms $\vec{m}_\text{t}=\text{(Re,Im)}[e^{i(K-K')\cdot R}]$
with
$\vec{\tau}_{x,y}=\left(\tau^x , \tau^y\right)$ are trivial mass terms that mix the two Dirac cones while preserving TR. 

The $t_2$ term is the TR breaking topological mass term $\sigma^z\tau^z$ familiar from the Haldane honeycomb model  \cite{haldane_model_1988}. Thanks to the $\sqrt{3}$ factor this term dominates over the trivial mass term leading to non-trivial topology with Chern number $C{=}\mu^z{=}\pm 1$.
This relation between the flux on each disclination in the SW and the sign of its contributed chirality $C$ is consistent with that found for 5-7 dislocations \cite{borhani_realspace_2025}, though such relations may not be universal  \cite{neehus_genuine_2025,xu_chirality_2025,seth_generation_2025}.
Summing over multiple scattering events to all orders in the T-matrix formalism (see End Matter)  renormalizes the topological mass term to $\sqrt{3}t_2\rightarrow \frac{18\sqrt{3}\pi t_2
}{(4\pi-3\sqrt{3})(t_1^2+t_2^2)+18\pi-12\pi t_1}$ 
giving a significantly enhanced $t_1{=}t_2{=}1$  gap of $ 11.6/(2 \mathcal{N}_c)$ for one defect per $2 \mathcal{N}_c$ sites. %
Each defect thus serves as a topological mass term with sign set by the dynamically fluctuating $\mu^z_r$.

\begin{figure}[t]
    \centering
    \includegraphics[width=\columnwidth]{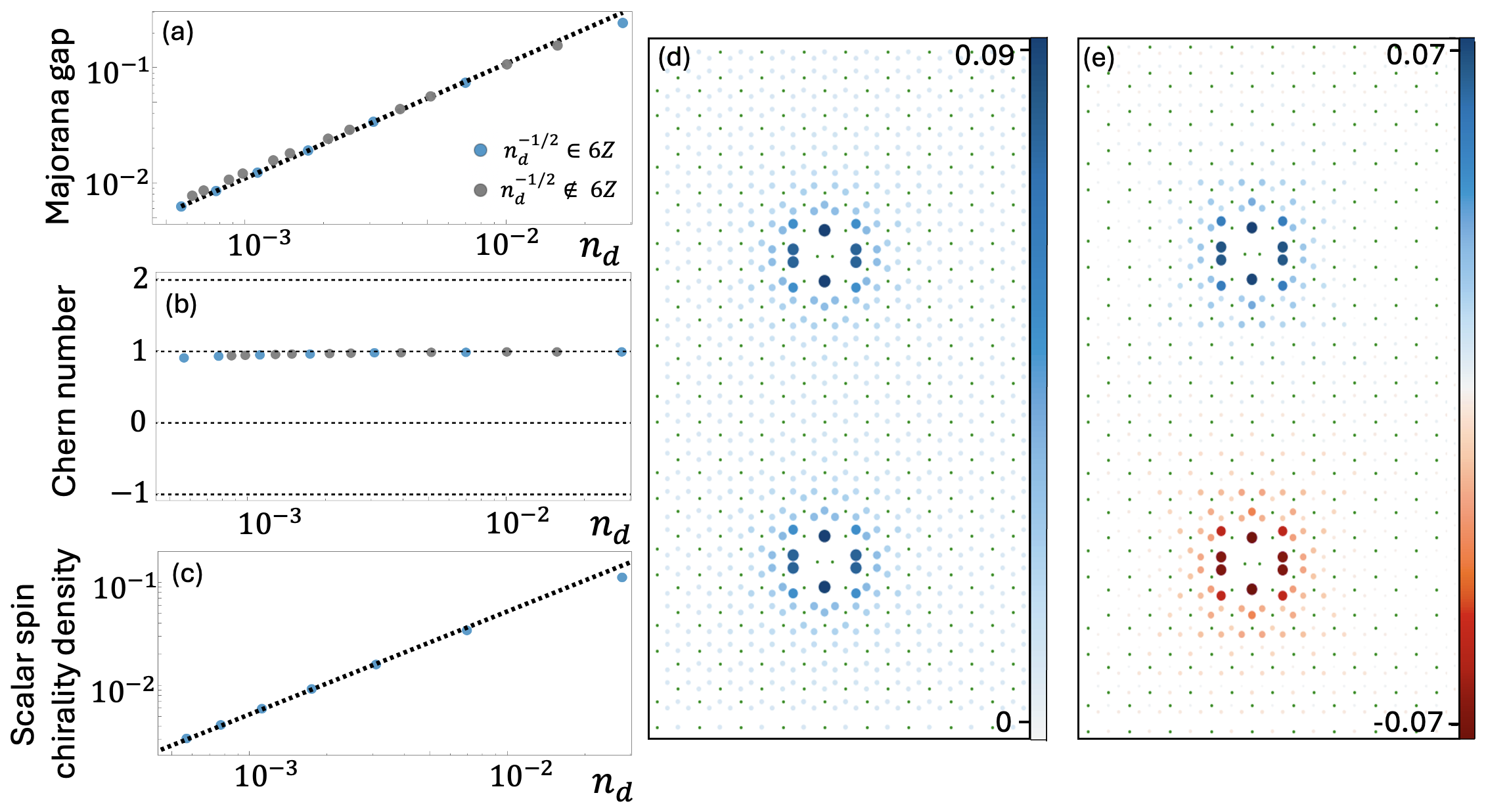}
    \caption{{\bf Chirality generation from local lattice defects.} 
    (a) 
    A density $n_d$ superlattice of $\mu^z_r=1$ defects     
    produces a finite Majorana gap ($\approx 10.8 \ n_d$, dotted line). (b) This gap is topological, giving  Majorana fermions a quantized Chern number $C=1$. (c,d,e) The resulting chirality is observable in %
    scalar spin chirality (SSC) from individual defects. (c) Average SSC density per site ($\approx 5.3 n_d$, dotted line). %
    (d,e) SSC of each 3-spin trio (plotted at NNN bonds of honeycomb sites [green dots]) for two SW defects with $\mu^z$ aligned (d) and antialigned (e). Inner core SSC %
    omitted for clarity. 
    }
    \label{fig2}
\end{figure}

This T-matrix result can be verified numerically for finite defect densities $n_d$.   As shown in Fig.~\ref{fig2}(a) and (b),  for arrays of $\mu^z{=}1$ Lieb-flux defects with density $n_d$ (number of defects divided by number of sites), the Majorana spectrum gains a gap $\Delta \approx 10.8 n_d$, similar to the T-matrix result. Indeed this gap always produces Chern number $C{=}1$; no $C\geq 2$ is observed.
To extend this result to random defects we consider periodic boundary systems with randomly placed defects and compute the Bott index.
\cite{loring_disordered_2011,hastings_almost_2010}
When all defects have $\mu^z=1$ ($\mu^z=-1$), the Bott index is found to be $B=1$ ($B=-1$). 
These results imply that finite density configurations of aligned-$\mu^z$ defects will generate the Non-Abelian chiral QSL at $T=0$ at arbitrarily small densities.

\paragraph{\textit{Scalar spin chirality and orbital magnetization --- }}

The topological Majorana Chern number also manifests in experimentally observable probes related to a local order parameter.
To quantify such a real-space contribution of each defect to the system's chirality we use the local Chern marker (see Ref.~\cite{seth_generation_2025} Sec.~VI) as well as the 
scalar spin chirality (SSC) defined by the triple product of three spins around a triangle, 
$
    \hat{\chi}_{ijk}=\sigma_i\cdot \left(\sigma_j\times \sigma_k\right)
$. 
Here $\sigma$ are spins on three sites $i,j,k$  listed in counterclockwise order. For the Kitaev model the expectation value of SSC in a fixed flux sector can be computed in terms of Majorana fermions as
$
    \langle\hat{\chi}_{ijk}\rangle=i \langle  u_{ij} u_{jk} c_ic_k\rangle
$. 
As shown in Fig.~\ref{fig2}(c,d,e), even though individual spins have exactly zero expectation value, in the  presence of SW defects the SSC gains  contributions proportional to  $\mu^z_r$, consistent with prior results for polycrystalline defects \cite{grushin_amorphous_2023}. 
These contributions are amplified for aligned $\mu^z$, generating a large average SSC per site of $\approx 5.3 n_d$ at small $n_d$, with much larger densities near each defect. %
Unlike the case of a field-induced chiral QSL \cite{kitaev_anyons_2006} here SSC arises at zero external field. Even in an applied field $B$ the defect-induced SSC can dominate ($\mathcal{O}( n_d)$ vs.\ $\mathcal{O}( (B/J_K)^3)$) with additional peaks near defects.

The scalar spin chirality generated by SW defects is experimentally observable both directly and via conventional electron orbital magnetization. 
Electronic charge fluctuations in the Mott insulator control the coupling between SSC and electron orbital moments. \cite{shindou_orbital_2001, motrunich_orbital_2006, bulaevskii_electronic_2008} If the Kitaev model arises from a parent extended Hubbard model with Hubbard $U$ and hopping $t$, the electronic orbital magnetization includes a term $\chi t^3/U^2$. 
This orbital magnetization can also be observed in finite frequency AC optical Hall conductivity via Faraday and Kerr rotations. 
\footnote{We also note that if the system is doped with itinerant electrons then SSC also produces Hall conductivity at zero frequency, through multiple mechanisms, including the topological Hall effect well known in skyrmion lattices \cite{kurumaji_skyrmion_2019} as well as anomalous Hall effect induced by spin chirality skew scattering \cite{ishizuka_spin_2018}. }

\paragraph{\textit{Emergent ferromagnetic long-range interactions --- } }

Having observed that a finite density of aligned $\mu^z=1$ defects generates SSC and opens a topological gap producing Chern number $C=1$, we now turn to the question of the emergent interactions that would tend to align or antialign $\mu^z_r$ of distant defects. 
This will also determine the  temperature of the chiral spin liquid instability.

\begin{figure}
    \centering
    \includegraphics[width=1\linewidth]{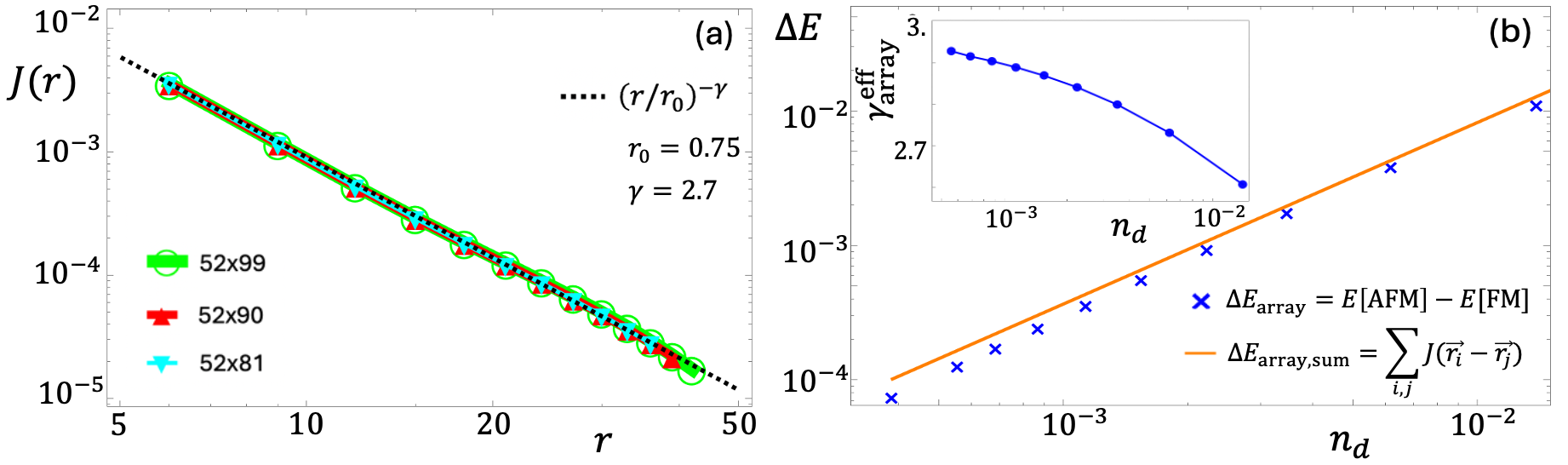}
    \caption{{\bf Emergent long-range interaction between defect chiralities.}  (a) 
    Energy differences $\Delta E$ between aligned and antialigned $\mu^z$ for two SW defects separated by $\vec{r}$ in finite PBC systems of linear length $L$ can be modeled as $\Delta E/2 = J(r)+J(L-r)+J_0$, with $r{=}|\vec{r}|,L$ measured in real space (NN bond length 1). The data ($\Delta E/2{-}J(L{-}r){-}J_0)$ for $L_x {\times} L_y$ systems
    is well fitted by $J(r) = (r_0/r)^\gamma$ with $r_0=0.75$ and $\gamma=2.7$ (dotted line). 
    (b) To verify the long range form of $J(r)$ we numerically computed the energy difference $\Delta E_\text{array}$ between ferromagnetic  and  N\'eel antiferromagnetic $\mu^z_r$ configurations of infinite defect arrays with density $n_d$ (blue crosses). The data shows remarkable agreement with an analytical infinite sum of the approximated  $J=(r_0/r)^\gamma$ (orange line). The exponent $\gamma\approx 2.7$ is renormalized from $\gamma_0=3$ by finite $n_d$,  as visible from the $n_d$ array data (inset). 
    See Ref.~\cite{seth_generation_2025} Sec.~VII for discussions of $\Delta E$, the $J(r)$ array sum, $J(\vec{r})$ anisotropies, generation of $J_0$ by PBC, as well as a third independent method of extracting $J(r)$ which shows similar results.  
    }
    \label{fig3}
\end{figure}

Consider two defects whose flux configurations $\mu^z_r$ are either the same or opposite, corresponding to Fig.~\ref{fig2}(d,e) respectively. The large difference in net SSC suggests that the energies of the two configurations will also differ, and hints that the configuration with larger net SSC may be energetically preferred. 
We computed this energy difference $\Delta E$ between aligned and anti-aligned defect chiralities $\mu^z$ for various pairs of defects and with either periodic or open boundary conditions.
Within the range of separations consistent with the relevant range of densities $n_d\approx 10^{-4}$--$10^{-2}$ we find that $\Delta E$ is well approximated as a simple function of the inter defect real space distance $r-r'$.
We describe this behavior in terms of an emergent Ising interaction $J(r-r')$ with  $\Delta E = 2J$,
\begin{align}
    &H^{\text{SW}}_{\text{Ising}}=-\frac{1}{2}\sum_{ r, r'}{J}\left(\vec{r}-\vec{r}'\right)\mu^z_{ r}\mu^z_{ r'},\quad J(\vec{r})\approx \left(\frac{r_0}{r}\right)^{\gamma}
    \label{eq_sw_int}
\end{align}
Aligned (i.e.\ matching) $\mu^z_r$ produce lower energy, implying $J$ is ferromagnetic in $\mu^z$.
As shown in Fig.~\ref{fig3}, the approximately power law form of $J(r)$ is found to give an excellent fit to the numerical data. 
The exponent $\gamma\approx 2.7$ becomes equal to exactly $\gamma_0= 3$ for continuum Dirac fermions in the limit of dilute defects $n_d\rightarrow 0$  where it controls the RKKY interactions \cite{kogan_rkky_2011}. The small Dirac mass of order $n_d$ generates a decay length which is much larger than the typical defect separation  $n_d^{-1/2}$, enabling the gapless power law form to persist. The nonzero $n_d$ can then suppress 
\footnote{The observation that $\gamma$ is suppressed rather than enhanced can be interpreted by considering the graphene $r^{-3}$ RKKY as arising from the energy or equivalently $k$ integral of the $r^{-2}$ spin LDOS for energies near the Dirac point \cite{liu_magnetic_2009}. %
Smoothly cutting off this $k$ integral by $n_d^{1/2}$ can produce an approximate $r^{-\gamma}$ form with $2<\gamma<3$.}
the effective $\gamma$, producing a longer range interaction, as shown in Fig.~\ref{fig3}(b).

\paragraph{\textit{Finite temperature instability --- }}

This  long range emergent ferromagnetic interaction 
between the fluctuating $\mu^z_r$ mass terms
produces dramatic effects. 
Finite densities of SW defects drive a phase transition of spontaneous time-reversal symmetry breaking associated with aligned  $\mu^z_r$  chiralities, leading to a chiral spin liquid ground state. %
The symmetry breaking transition  has an order parameter given by net chirality and measurable by SSC or electronic orbital magnetization. In other words, the fluctuating spins produce nonzero average expectation value for SSC at temperatures $T<T_c$. The ensuing chiral QSL phase also exhibits fractionaliztion, which as always in 2D is sharply defined in terms of Wilson loops only at $T=0$ but nevertheless has visible consequences at finite $T$. In the present case one such consequence is the chirality associated with lattice defects.  Note that the $Z_2 \times Z_2$ spin rotation symmetry of Kitaev Hamiltonians is  sufficient to distinguish this phase from conventional spin ferromagnets with $\langle \vec{S} \rangle \neq 0$. The present case produces orbital, but not spin, electronic magnetization. 

The transition temperature $T_c$ can be estimated from $J(r)$ via the usual mean field theory controlled by the long range of $J(r)$ (see End Matter), as
\begin{align}
 T_c  =8\sqrt{3}\pi
 \frac{(r_0/3a)^\gamma}{\gamma-2}  n_d J_K
 \approx 2 n_d J_K
\end{align}
where the coefficient 2 of $2 n_d J_K$ has uncertainty $\pm 1$ across approximations and variability in $J(r)$. 

The critical $T_c$ is sensitive to the value of $\gamma$. As discussed in Ref.~\cite{seth_generation_2025} Sec.~IX, Hamiltonian perturbations $\delta H_r$ arising near each defect can modify the effective $\gamma$.  The computed $T_c$ rises to $10 n_d$ for $\gamma \approx 2.3$ before diverging when $\gamma \rightarrow 2$. In that limit we expect $T_c$ to be cut off by the SW flux gap, $\approx 0.08 J_K$, independent of $n_d$.

Interestingly, the long range interaction
$r^{-\gamma}$ satisfying $2d/(\gamma-d)\geq 4$ for all $2<\gamma \leq 3$ in this $d=2$ model implies that the 
critical behavior of the Ising transition at $T_c$ is described by the Gaussian fixed point, \cite{fisher_critical_1972,aizenman_critical_1988} with critical exponents taking the exact mean field values
$\nu_\text{crit}=1/(\gamma-2)$,
$\beta_\text{crit}=1/2$ and
$\gamma_\text{crit}=1$.

\paragraph{\textit{Discussion --- }}
In this work we showed how the presence of a certain class of local crystallographic defects generates an instability from the gapless Kitaev QSL to the non-Abelian chiral QSL.
The phase transition arises from Majorana fermion Dirac cones coupled to the fluctuating $\mu^z_r$ fluxes. These  $\mu^z_r$ act as local mass terms producing localized chirality. Their coupling to the Dirac cones also provides the $\mu^z_r$ with an emergent long range interaction. 

At finite defect densities the emergent interaction leads to ferromagnetic ordering of the defect chiralities via a finite temperature phase transition. 
The transition temperature $T_c$ is set by the spin exchange energy $J_K$ and the defect density, and can be further enhanced by tuning the emergent long range interactions that produce it. For example, for $J_K$ values of several meV (as in RuCl$_3$) \cite{savary_quantum_2016,
hermanns_physics_2018, takagi_concept_2019, trebst_kitaev_2022,  matsuda_kitaev_2025}, $T_c\approx  1$ Kelvin  is achievable even for defect densities as low as one defect per hundreds of sites.
As usual for 2D systems the spin liquid's fractionalization  then becomes sharply defined at $T=0$, producing a non-Abelian chiral spin liquid phase.

Experimentally, the emergent chiral QSL can be detected in various ways. The well known signatures of the usual field-induced Kitaev chiral QSL phase apply, from the half integer quantized thermal Hall effect \cite{kitaev_anyons_2006,yokoi_halfinteger_2021} to recent proposals including low energy edge signatures \cite{zhang_lowenergy_2025,zhang_probing_2025}. In the present case the emergent chiral QSL can also be detected through the electronic orbital magnetization via e.g.\ Faraday or Kerr rotation as discussed above, which occurs with zero external magnetic field and zero spin magnetization. 
We also note recent work proposing Raman circular dichroism \cite{koller_raman_2025, koller_spinlattice_2025}. 
Experimentalists may also use local magnetization probes
\cite{vasyukov_scanning_2013,finkler_selfaligned_2010,mclaughlin_local_2023,du_control_2017,grover_chern_2022}
to observe the defect induced chiral QSL puddles that may form due to defect clusters even above $T_c$. Similar ``Chern mosaics'' have already been observed by probing orbital magnetism using a scanning SQUID on a tip \cite{grover_chern_2022}. 

Our results raise the possibility that candidate Kitaev materials %
whose actual ground state shows scalar spin chirality and electronic orbital magnetization could instead be realizing the defect-induced chiral QSL phase. In such a case, a scanning probe measuring an inhomogeneous electronic orbital magnetization or scalar spin chirality, which is further peaked near nonmagnetic lattice defects, would serve as a disorder-based signature of fractionalized Majorana fermions.

\paragraph{\textit{Acknowledgments.}}

This work was supported by the U.S. Department of Energy, Office of Science, Basic Energy Sciences, under Early Career Award Number DE-SC0025478. We thank Natalia Perkins,  Masaki Oshikawa and Gabor Halasz  for helpful discussions.

The data  that support the findings of this article are openly available at the GT Digital Repository \cite{kimchi_georgia_2026}.

\bibliography{references}

\begin{appendix}

\section{End Matter}

\paragraph{\textit{\label{appen_tmatrix}Appendix A: Scattering T-matrix analysis  ---  }}

The locality of the Lieb-flux SW term 
$V^\text{SW}_\text{Lieb}$ (Eq.~\ref{eq_sw_real space}) enables an analysis in terms of the scattering T-matrix formalism, which takes into account  the  scattering of the itinerant Majoranas  from the defect at all orders. Viewing $V^\text{SW}_\text{Lieb}$  as a localized (but non-onsite) impurity ``potential'' $V$, its T-matrix is given by
\begin{align}
    T(E)=V (1-G_0(E) V)^{-1}.
\end{align}
Here $G_0(E)$ is the Green's function of the unperturbed Hamiltonian at energy $E$.  

In the present case, the  Green's function of the unperturbed  zero-flux clean honeycomb Kitaev model $H_0$ is easily obtained by mapping the $c$ Majorana $H_0$ model to an equivalent model of complex fermions $f$ hopping on the honeycomb lattice with imaginary amplitudes (oriented from sublattice $A$ to $B$). 
This mapping, which double counts degrees of freedom due to $f$ being complex, can be done by replacing $A_{i,j} c_i c_j$ by $A_{i,j} f^\dagger_i f_j + \text{H.c}$;  see Ref.~\cite{seth_generation_2025} Appendix A. 
This imaginary-hopping $H_0$ is gauge equivalent to the familiar graphene model of real hopping on the honeycomb. The gauge transformation modifies $B$ sublattice operators by $f_B\rightarrow i f_B$, equivalent (up to an irrelevant overall phase) to a rotation of  $\vec{\sigma}$ around $\sigma^z$ by  angle $-\pi/2$. For $P[H_0]$  this yields the familiar Dirac cone expression
 $P[H_0] \rightarrow -v_F\sum_{ q} {\psi}^{\dagger}_{ q}\left(q_x\tau^z\sigma^x+ q_y\sigma^y\right){\psi}_{ q}$.

 Taking the well known lattice Green's function of graphene \cite{kot_band_2020} and 
 applying the inverse gauge transformation 
 $f_B\rightarrow -i f_B$ yields the $G_0(E)$ corresponding to imaginary hoppings, which can be combined with $V^\text{SW}_\text{Lieb}$  of Eq.~\ref{eq_sw_real space} to form the T-matrix in real space.
  Since we are concerned with identifying gap openings around $E=0$, it is sufficient to take $G_0$ at  zero energy $G_0(E=0)$. 
  The last step is to project $T(E=0)$ to  the low energy Dirac cone theory $P$. 
The computation can alternatively be performed in the real hopping gauge; see Ref.~\cite{seth_generation_2025} Appendix C, which also includes results for the T-matrices of other flux configurations.

Let us start by discussing the first order term (in an expansion in defect strength $t_{1,2}$) which is just $V$. Projecting to the low energy theory we find (as in Eq.~\ref{eq_sw perturbation} but with suppressed terms now restored)
\begin{align}
    &P[V_\text{Lieb}^\text{SW}]=\frac{1}{\mathcal{N}_c}\psi^\dagger_q\left(\sqrt{3}\mu^zt_2 \tau^z\sigma^z-t_1 \vec{m}_\text{t}\cdot(\tau^x, \tau^y)\sigma^y\right.\nonumber\\
    &\hspace{4cm}\left.-2t_1\vec{\phi}\cdot (\sigma^y,\sigma^x\tau^z)\right)\psi_q
\end{align}
where 
$\vec{\phi}{=}(\text{Re , Im})[e^{i \pi/3}]$ and
$\vec{m}_t{=}(\text{Re , Im})[e^{i (K-K')\cdot R}]$.
The  terms with coefficient $\vec{\phi}$ merely shift the position of Dirac cones in an inversion symmetric manner, and do not open a gap or break TR.

Higher order scattering events renormalize this first order result. Their exact resummation for a single defect is encompassed by the $(1-G_0(E) V)^{-1}$ denominator of the T-matrix. Computing 
$T(E{=}0)$ and projecting to low energy we find the same terms but with renormalized coefficients: 
\begin{align}
    & P[T_\text{Lieb}^\text{SW}]=\frac{1}{\mathcal{N}_c}f_1 \psi_q^\dagger\left(a_1 t_2\mu^z\tau^z\sigma^z-b_1 \vec{m}_t\cdot \left(\tau^x,\tau^y\right)\sigma^y\right.\nonumber\\
    &\hspace{4cm}\left.-c_1\sigma^y-d_1\sigma^x\tau^z
    \right)\psi_q
    \label{eq_tmatrix_lieb}
\end{align}
where ${f}_1= \left((4\pi-3\sqrt{3})(t_1^2+t_2^2)+18\pi-12\pi t_1\right)^{-1}$,~~ $a_1= 18\sqrt{3}\pi$,~~ $b_1=18\pi t_1-(9\sqrt{3}+6\pi)(t_1^2+t_2^2)$,~~
    $c_1= 18\pi t_1+(3\pi-9\sqrt{3})(t_1^2+t_2^2)$,~~ $d_1= -9\sqrt{3}\pi(t_1^2+t_2^2-2t_1)$ .
    The chiral $\mu^z$ (i.e.\ first) term still comes with an overall coefficient of $t_2$. 

Comparing the first and second terms, which give the competing topological and trivial mass terms, we find that at $t_1=t_2=1$ the renormalization significantly enhances the topological term  $a_1$  relative to the trivial term $b_1$. Thus the resummation over scattering events provides additional protection for the nontrivial $C=\pm 1$ Chern numbers.

\paragraph{\textit{Appendix B: Random ferromagnetic Ising model  --- }}
\label{appen_ising}

To estimate the transition temperature of a random site FM Ising model $H_{\mu^z}$ it is sufficient to use mean field theory since the interaction is long ranged. (At $\gamma{=}0$ mean field is exact.)
Let
\(H_{\mu^z} = -(1/2) \sum_{i,j} J_{i,j} \mu^z_i \mu^z_j\). The factor 1/2 occurs since each bond is double counted. 
The mean field $H_{\mu^z}$ becomes
\(m \sum_{i,j} J_{i,j} \mu_j^z \) giving magnetization 
\( m = \tanh  [(m/T) (1/N)\sum_{i,j} J_{i,j}]\). 
Nonzero self consistent solutions appear for $T$ below 
\(T_c = (1/N)\sum_{i,j} J_{i,j}\).
This generalizes the familiar expression for n.n.\ coupling with coordination number $z$ where $H=-J\sum_{<ij>} \mu^z_i \mu^z_j$ and $T_c = z J$.

We assume an interaction $J(r) = (r/r_0)^{-\gamma}$ in units of $J_K$. 
Recall the distribution of distances \( r \) between any two randomly chosen points in a 2D area $A$ is $P(r) = 2 \pi r/A$. 
We compute mean field $T_c$ for random defects in a continuum limit as
\begin{align}
T_c &= (1/N)\sum_{i,j} J_{i,j}
= \sum_{j} J(r_{j})
= N \int dr P(r) J(r)
\nonumber\\ &
=  \frac{\pi r_m^2}{A_s}  \frac{r_0^\gamma}{r_m^\gamma} \frac{2}{\gamma-2} n_d 
= 
 8\sqrt{3}\pi
 \frac{(r_0/3a)^\gamma}{\gamma-2}  n_d 
\end{align}
where $r_m=3 a$ is the minimum defect separation and $A/N_s=A_s=3\sqrt{3}a^2/4$ is the area per site of the honeycomb lattice with $a$ the n.n.\ bond length.

\end{appendix}

\clearpage

\end{document}